\documentclass[aps,prb,reprint,groupedaddress]{revtex4-2}

\usepackage{float}
\usepackage{mathrsfs}
\usepackage{graphicx,amsfonts,bm,amssymb,amsmath,braket,hyperref}
\hypersetup{colorlinks=true,allcolors=[rgb]{0,0,1}}
\usepackage{mathptmx}

\begin{document}


\title{Phonon Stability and Sound Velocity of Quantum Droplets in a Boson Mixture}

\author{Qi Gu, Lan Yin}
\email[]{yinlan@pku.edu.cn}
\affiliation{School of Physics, Peking University, Beijing 100871, China}


\date{\today}

\begin{abstract}
Quantum droplets have been realized in experiments on binary boson mixtures and dipolar Bose gases.  In these systems, the mean-field energy of the Bose-Einstein condensation is attractive, and the repulsive Lee-Huang-Yang energy is crucial for stability.  However, since the Lee-Huang-Yang term is not taken into account when constructing the Bogoliubov Hamiltonian, in the droplet regime one faces the problem of dynamically unstable phonon modes.  In this work, we go beyond the Bogoliubov theory to study how the phonon mode is stabilized in the quantum droplet of a binary boson mixture.  Similar to Beliaev's approach to a single-component Bose gas, we compute higher-order contributions to the self-energy of the boson propagator.  We find that the interaction between spin and phonon excitations is the key for the phonon stability.  We obtain the sound velocity which can be tested by measuring the superfluid critical velocity of the droplet in experiments.  Beliaev damping of this quantum droplet is also discussed.
\end{abstract}


\maketitle



\textit{Introduction}--
Quantum droplet states of ultracold atoms are self-bound and can survive in the vacuum for considerable long time without trapping.  In a seminal work \cite{petrovQuantumMechanicalStabilization2015}, Petrov pointed out that attractive mean-field and repulsive Lee-Huang-Yang (LHY) energies \cite{leeEigenvaluesEigenfunctionsBose1957}  are the keys to the formation of quantum droplets in Bose gases.  In recent several years, quantum droplets have been successfully realized in experiments on dipolar Bose gases of $^{164}\mathrm{Dy}$ \cite{kadauObservingRosensweigInstability2016,ferrier-barbutObservationQuantumDroplets2016,ferrier-barbutLiquidQuantumDroplets2016,schmittSelfboundDropletsDilute2016,wenzelStripedStatesManybody2017}, $^{168}\mathrm{Er}$ \cite{chomazQuantumFluctuationDrivenCrossoverDilute2016} atoms, the homonuclear mixture of $ ^{39}\mathrm{K} $ \cite{cabreraQuantumLiquidDroplets2018,cheineyBrightSolitonQuantum2018,semeghiniSelfBoundQuantumDroplets2018}, and the heteronuclear $^{39}\mathrm{K}$-$^{87}\mathrm{Rb}$ mixture \cite{derricoObservationQuantumDroplets2019}.

The quantum droplet can be studied from many aspects, such as finite size effects \cite{cabreraQuantumLiquidDroplets2018,semeghiniSelfBoundQuantumDroplets2018,schmittSelfboundDropletsDilute2016}, low-dimension matter waves \cite{lepoutreProductionStronglyBound2016,cheineyBrightSolitonQuantum2018}, and supersolid properties \cite{tanziSupersolidSymmetryBreaking2019,guoLowenergyGoldstoneMode2019}. The current theoretical approaches are mainly based on extended Gross-Pitaevskii equation \cite{wachtlerGroundstatePropertiesElementary2016,baillieSelfboundDipolarDroplet2016,petrovUltradiluteLowDimensionalLiquids2016,wachtlerQuantumFilamentsDipolar2016,kartashovMetastabilityQuantumDroplet2019} or quantum Monte Carlo techniques \cite{saitoPathIntegralMonteCarlo2016,maciaDropletsTrappedQuantum2016,cintiSuperfluidFilamentsDipolar2017,bottcherDiluteDipolarQuantum2019}, but
the microscopic theory is still incomplete.  In the Bogoliubov theory of the quantum droplet \cite{petrovQuantumMechanicalStabilization2015}, the phonon excitation is unstable in long-wavelength limit.  It was postulated that the phonon excitations can be stabilized by integrating out higher-energy excitations, but never demonstrated so far.  In the computation of LHY energy, this unstable mode is ignored based on the argument that its contribution is negligible compared to that from the higher-energy mode.

In this work, we study the stability of the phonon mode in the quantum droplet of a dilute binary boson mixture. The method we use is developed from the Beliaev theory \cite{beliaevApplicationQuantumField1958,beliaevEnergySpectrumNonideal1958} which was also adopted in the study of renormalization to the magnon mass in a binary Bose gas \cite{utesovEffectiveInteractionsQuantum2018}. We go beyond the Bogoliubov approximation and compute the contribution to the boson self-energy from higher order fluctuations.  We find that in the dilute limit, the leading correction to the phonon energy comes from the interaction between phonon and spin excitations.  This correction is positive and larger in magnitude than the phonon energy from the Bogoliubov theory in the long wavelength limit.  Thus the phonon mode is stabilized by the interaction with the spin excitations.  We obtain the sound velocity and find it is consistent with the hypothesis of superfluid hydrodynamics of the quantum droplet.  This result can be tested in the experiment on the critical superfluid velocity of the quantum droplet.  Beliaev damping of the phonon mode is also discussed.

\textit{Model}--
We consider a binary Bose mixture described by the Hamiltonian $ H=H_0+H_{\text{int}} $. The single-particle term is given by
\begin{equation*}
H_0=\sum_\mathbf{p}\epsilon^0_{\mathbf{p}}(\hat{\alpha}^\dagger_\mathbf{p}\hat{\alpha}_\mathbf{p}+\hat{\beta}^\dagger_\mathbf{p}\hat{\beta}_\mathbf{p}),
\end{equation*}
where $ \hat{\alpha} $ and $ \hat{\beta} $ are the annihilation operators of the two components $ \alpha $ and $ \beta $, $ \epsilon^0_{\mathbf{p}}=\hbar^2\mathbf{p}^2/(2m) $, and $m$ is the mass of a boson. The $ s $-wave interaction term is given by
\begin{align*}
\nonumber
H_{\text{int}}&=\frac{1}{2V}\sum_{\mathbf{p}_1,\mathbf{p}_2,\mathbf{p}_3,\mathbf{p}_4}'\big(g_{\alpha\alpha}\hat{\alpha}^\dagger_{\mathbf{p}_1}\hat{\alpha}^\dagger_{\mathbf{p}_2}\hat{\alpha}_{\mathbf{p}_3}\hat{\alpha}_{\mathbf{p}_4} \\ &+g_{\beta\beta}\hat{\beta}^\dagger_{\mathbf{p}_1}\hat{\beta}^\dagger_{\mathbf{p}_2}\hat{\beta}_{\mathbf{p}_3}\hat{\beta}_{\mathbf{p}_4} + 2g_{\alpha\beta}\hat{\alpha}^\dagger_{\mathbf{p}_1}\hat{\beta}^\dagger_{\mathbf{p}_2}\hat{\beta}_{\mathbf{p}_3}\hat{\alpha}_{\mathbf{p}_4} \big),
\end{align*}
where $ V $ is the volume, and the total momentum is conserved, $ \mathbf{p}_1+\mathbf{p}_2=\mathbf{p}_3+\mathbf{p}_4 $.  The $s$-wave coupling constants are given by $ g_{ij}=4\pi\hbar^2 a_{ij}/m $, where $i,j=\alpha,\beta$, and $a_{ij} $ is the scattering length. In experiments, the low density quantum droplets are formed near the mean-field unstable-miscible boundary, with the density ratio of the two components fixed at $ n_\alpha/n_\beta=\sqrt{a_\beta/a_\alpha} $ \cite{petrovQuantumMechanicalStabilization2015}. Two characteristic length scales can be defined, $ \delta a=\eta(\sqrt{a_{\alpha\alpha}a_{\beta\beta}}+a_{\alpha\beta})/2 $ and $ a'=\eta(a_{\alpha\alpha}+a_{\beta\beta}-2a_{\alpha\beta})/4 $, where $ \eta=4n_\alpha n_\beta/(n_\alpha+n_\beta)^2 $ describing the density imbalance. The corresponding coupling parameters are defined as $ \delta g=4\pi\hbar^2 \delta a/m $ and $ g'=4\pi\hbar^2 a'/m $. In current experiments, the droplet is formed in the region with $a_{ii}>0$, $\delta a<0$, $\gamma=|\delta a|/a'  \ll 1$ in the dilute regime $\sqrt{n_ta'^3}\ll 1$ where $n_t$ is the total boson density.

In the mean-field approximation, the bosons in the ground state are condensed with the mean-field energy density given by $ \dfrac{1}{2}\sum_{ij}g_{ij}n_{i}n_{j} $, which is negative for the quantum droplet.  LHY energy density can be obtained in the Bogoliubov approximation, $\dfrac{1}{2}\sum_{\pm}\sum_\mathbf{p}\{\mathcal{E}_{\pm,\mathbf{p}}-\mathcal{E}_{\pm,\mathbf{p}}^2/(2\epsilon^0_\mathbf{p})-\epsilon^0_\mathbf{p}/2\}$, where $\mathcal{E}_{\pm,\mathbf{p}}$ are magnetic and phonon excitation energies \cite{larsenBinaryMixturesDilute1963}.  The phonon energy in the long wavelength limit is given by
\begin{equation*}
\dfrac{\hbar p}{\sqrt{2m}}\sqrt{\sqrt{g_{\alpha\alpha}g_{\beta\beta}}n_t - \sqrt{g_{\alpha\alpha}g_{\beta\beta}n_t^2+4n_{\alpha} n_{\beta}(g_{\alpha\beta}^2-g_{\alpha\alpha}g_{\beta\beta})}},
\end{equation*}
which is imaginary for the quantum droplet, indicating dynamic instability of the system, but the phonon contribution to LYH energy is much smaller in magnitude than that from magnetic excitations.  In Ref.~\cite{petrovQuantumMechanicalStabilization2015}, it was postulated that the phonon mode can be stabilized by integrating out high-energy excitations and its contribution to LHY energy can be ignored.  The total energy density of the quantum droplet is thus given by \cite{petrovQuantumMechanicalStabilization2015}
\begin{equation}\label{Total_Engergy}
-\dfrac{\gamma'}{2}\sqrt{g_{\alpha\alpha}g_{\beta\beta}}n_t^2 + \dfrac{64}{15\sqrt{\pi}}\sqrt{g_{\alpha\alpha}g_{\beta\beta}}n_t^2\sqrt{n_t (a_{\alpha\alpha}a_{\beta\beta})^{3/2}},
\end{equation}
and a positive compressibility can be obtained from its second derivative with density, where
$ \gamma'=-2(\sqrt{a_{\alpha\alpha}a_{\beta\beta}}+a_{\alpha\beta})/(\sqrt{a_{\alpha\alpha}}+\sqrt{a_{\beta\beta}})^2 $. The quantum droplet is self-bound in vacuum with zero pressure, satisfying the condition
$\gamma'=\dfrac{64}{5\sqrt{\pi}}\sqrt{n_t (a_{\alpha\alpha}a_{\beta\beta})^{3/2}} $,
from which the droplet density can be obtained \cite{petrovQuantumMechanicalStabilization2015}, $ n_t=\dfrac{25\pi\gamma'^2}{4096}(a_{\alpha\alpha}a_{\beta\beta})^{-3/2} $. In the dilute region, the denity $n_t$ is the same as the total condensate density $n$ to the leading order of the gas parameter.  In current experiments, the quantum droplet is formed in region where the two parameters $\gamma$ and $\gamma'$ are of the same order, $\gamma/\gamma'=1-\gamma\approx 1 $.  In this region, the parameter $\gamma'$ can be replaced by $\gamma$ in Eq. (\ref{Total_Engergy}).  In the following we consider higher-order fluctuations beyond Bogoliubov approximation to study how the phonon mode is stabilized in the long wavelength limit in the experimental region.

\textit{Symmetric Case}--
For illustration purposes, we first consider the system with symmetric intraspecies interactions, $g_{\alpha\alpha}=g_{\beta\beta}$.  In this case, the density of each component in the condensate is equal, $n_\alpha=n_\beta$, i.e.  $ \eta=1 $.  For simplicity, we assuming the expectation values of the field operators are equal, $\langle \hat{\alpha}_0 \rangle =\langle \hat{\beta}_0 \rangle=\sqrt{nV/2}$, and introduce new operators related to density and spin fluctuations, $\hat{c}_\mathbf{p}=(\hat{\alpha}_\mathbf{p}+\hat{\beta}_\mathbf{p})/\sqrt{2} $ and $ \hat{d}_\mathbf{p}=(\hat{\alpha}_\mathbf{p}-\hat{\beta}_\mathbf{p})/\sqrt{2}$, where $\langle \hat{c}_0 \rangle =\sqrt{nV}$, $\langle \hat{d}_0 \rangle=0$, and $n$ is the total density of the condensate.  The Hamiltonian in this representation is given by
\begin{align}
\nonumber
H&=\sum_\mathbf{p}\epsilon^0_\mathbf{p}(\hat{c}^\dagger_\mathbf{p}\hat{c}_\mathbf{p}+\hat{d}^\dagger_\mathbf{p}\hat{d}_\mathbf{p}) +\frac{1}{2V}\sum_{\mathbf{p}_{1},\mathbf{p}_2,\mathbf{p}_3,\mathbf{p}_4}'\Big\{ \delta g\big[\hat{c}^\dagger_{\mathbf{p}_1}\hat{c}^\dagger_{\mathbf{p}_2}\hat{c}_{\mathbf{p}_3}\hat{c}_{\mathbf{p}_4}\\
\nonumber
&+\hat{d}^\dagger_{\mathbf{p}_1}\hat{d}^\dagger_{\mathbf{p}_2}\hat{d}_{\mathbf{p}_3}\hat{d}_{\mathbf{p}_4}
+(\hat{c}^\dagger_{\mathbf{p}_1}\hat{d}^\dagger_{\mathbf{p}_2}\hat{d}_{\mathbf{p}_3}\hat{c}_{\mathbf{p}_4}+\text{h.c.})\big]\\
&+g'\big[(\hat{c}^\dagger_{\mathbf{p}_1}\hat{c}^\dagger_{\mathbf{p}_2}\hat{d}_{\mathbf{p}_3}\hat{d}_{\mathbf{p}_4}+\text{h.c.})+(\hat{c}^\dagger_{\mathbf{p}_1}\hat{d}^\dagger_{\mathbf{p}_2}\hat{c}_{\mathbf{p}_3}\hat{d}_{\mathbf{p}_4}+\text{h.c.})\big]
\Big\},
\end{align}
where $ g'=(g_{\alpha\alpha}-g_{\alpha\beta})/2 $, $ \delta g=(g_{\alpha\alpha}+g_{\alpha\beta})/2 $, and the total momentum is conserved, $\mathbf{p}_{1}+\mathbf{p}_2=\mathbf{p}_3+\mathbf{p}_4$.
In the Bogoliubov approximation, the density and spin fluctuations are decoupled, and the Hamiltonian is quadratic and given by
\begin{align}
\nonumber
H_B=&\dfrac{\delta g n^2 V}{2} +\sum_{\mathbf{p}}(\epsilon^0_\mathbf{p}+2\delta g n)\hat{c}^\dagger_\mathbf{p}\hat{c}_\mathbf{p}+ \dfrac{\delta g n}{2}\sum_{\mathbf{p}\ne 0}(\hat{c}^\dagger_\mathbf{p}\hat{c}^\dagger_\mathbf{-p}+\text{h.c.})\\
+&\sum_{\mathbf{p}}(\epsilon^0_\mathbf{p}+\delta g n+ g'n)\hat{d}^\dagger_\mathbf{p}\hat{d}_\mathbf{p}+\dfrac{g'n}{2}\sum_{\mathbf{p}\ne 0}(\hat{d}^\dagger_\mathbf{p}\hat{d}^\dagger_\mathbf{-p}+\text{h.c.})
.
\end{align}
The corresponding thermodynamical potential in the grand canonical ensemble $ K_B=H_B-\mu^{(0)}\sum_\mathbf{p}(\hat{c}^\dagger_\mathbf{p}\hat{c}_\mathbf{p}+\hat{d}^\dagger_\mathbf{p}\hat{d}_\mathbf{p}) $ with mean-field chemical potential $ \mu^{(0)}=\delta gn $ can be diagonalized~\cite{Fetter}, leading to the spin excitation with energy $\mathcal{E}^d_\mathbf{p}=\sqrt{\epsilon^0_\mathbf{p}(\epsilon^0_\mathbf{p}+2g'n)}$ and the phonon excitation with $ \mathcal{E}'^c_\mathbf{p}=\sqrt{\epsilon^0_\mathbf{p}(\epsilon^0_\mathbf{p}+2\delta gn)} $.  For quantum droplets with $\delta g<0$, the phonon energy is imaginary in the Bogoliubov theory.

\begin{figure}[]
	\includegraphics[width=\columnwidth]{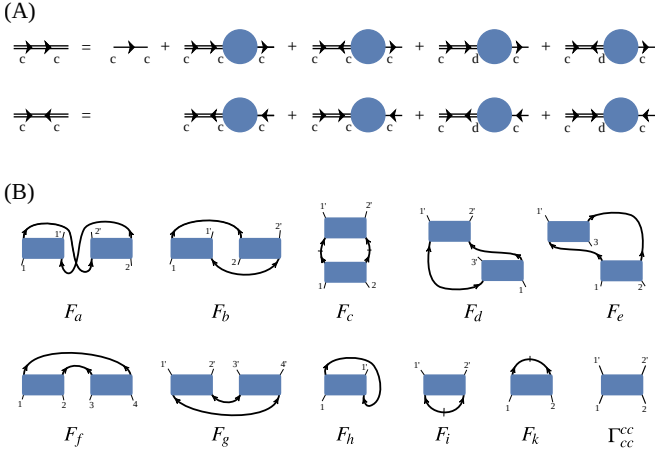}
	\caption{\label{fig1} Feynman diagrams of boson self-energies. (A) Dyson's equation. The single line with an arrow is the noninteracting Green's function. The double line with two same arrows is the normal part of the full Green's function, and that with two opposite arrows is the anomalous part.  The filled circles are self-energies. (B) Leading second-order diagrams of self-energies $ \Sigma_{cc} $. The external lines represent particle $ c $ in or out of condensed state. The internal lines (bold lines) linking two $ T $-matrices are the first-order Green's functions  $ G_d$ and $ \hat{G}_d$.~\cite{supplementary}}
\end{figure}

We study the phonon stability by seeking the effect of higher-order fluctuations.  The Green's function of the boson is a matrix, defined as
\begin{equation}
\mathbf{G}(\mathbf{p},t_1-t_2)=-i\braket{T\{\Psi_\mathbf{p}(t_1)\Psi_\mathbf{p}(t_2)^\dagger\}},
\end{equation}
where $ \Psi_\mathbf{p}(t)=[\hat{c}_\mathbf{p}(t),\hat{c}_\mathbf{-p}^\dagger(t),\hat{d}_\mathbf{p}(t),\hat{d}_\mathbf{-p}^\dagger(t)]^\intercal $, and $T$ is the time-ordering operator.  The Dyson's equation is given by
\begin{widetext}
\begin{equation}\label{Dyson}
\mathbf{G}^{-1}(p)=
\begin{pmatrix}
p^0+\mu-\epsilon^0_p-\Sigma_{cc}^{11}(p) & -\Sigma_{cc}^{20}(p) & -\Sigma_{cd}^{11}(p) & -\Sigma_{cd}^{20}(p)\\
-\Sigma_{cc}^{20}(p) & -p^0+\mu-\epsilon^0_p-\Sigma_{cc}^{11}(-p) & -\Sigma_{cd}^{20}(p) & -\Sigma_{cd}^{11}(-p)\\
-\Sigma_{cd}^{11}(p) & -\Sigma_{cd}^{20}(p) & p^0+\mu-\epsilon^0_p-\Sigma_{dd}^{11}(p) & -\Sigma_{dd}^{20}(p)\\
-\Sigma_{cd}^{20}(p) & -\Sigma_{cd}^{11}(-p) & -\Sigma_{dd}^{20}(p) & -p^0+\mu-\epsilon^0_p-\Sigma_{dd}^{11}(-p)
\end{pmatrix},
\end{equation}
\end{widetext}
where $\mu$ is the chemical potential, $p=(p_0, \mathbf{p})$, $p_0$ is the frequency, and
the non-interacting Green's function is given by $ G^0(p)=1/( p^0 +\mu -\epsilon^0_\mathbf{p}+i\delta) $. The proper self-energy is a block matrix,
\begin{equation*}
\Sigma=
\begin{pmatrix}
\Sigma_{dd} & \Sigma_{cd}\\
\Sigma_{cd} & \Sigma_{cc}
\end{pmatrix},
\end{equation*}
where each two-by-two block matrix $ \Sigma_{ij} $ is given by
\begin{equation*}
\Sigma_{ij}(p)=
\begin{pmatrix}
\Sigma_{ij}^{11}(p) & \Sigma_{ij}^{20}(p)\\
\Sigma_{ij}^{20}(p) & \Sigma_{ij}^{11}(-p)
\end{pmatrix}.
\end{equation*}
Following Beliaev's notation \cite{beliaevApplicationQuantumField1958,beliaevEnergySpectrumNonideal1958}, the superscript $11$ of $\Sigma_{ij}^{11}$ refers to an ingoing and an outgoing external line of
particle $ i $ and $ j $ in the Feynman diagram, and the superscript $20$ of $\Sigma_{ij}^{20}$ refers to two outgoing lines, as shown in Fig.~\ref{fig1}(A).
In the dilute region, the gas parameters $\delta \phi=\sqrt{n|\delta a|^3}$ and $\phi'=\sqrt{na'^3}$ are very small, serving as expansion parameters.  The first-order self-energy is consistent with Bogliubov theory, $ \Sigma_{cd}^{(1)}(p)=0 $, $ \Sigma_{cc}^{(1)}(p)$ and $\mu^{(1)}$ is of the order of $ \delta g n $, which are much smaller than
$ \Sigma_{dd}^{(1)}(p)$ in the region with $ |\delta a|\ll a' $. Thus higher-order contributions to the self-energy are crucial for phonon stability.  Nonetheless, the first-order spin excitations are stable, with the normal (diagonal) and anomalous (off-diagonal) Green's functions given by
\begin{equation}\label{G_d}
\begin{aligned}
G_d(p)&=(p^0+\epsilon^0_\mathbf{p}+g'n)/({p^0}^2-{\mathcal{E}^d_\mathbf{p}}^2+i\delta),\\
\hat{G}_d(p)&=-g'n/({p^0}^2-{\mathcal{E}^d_\mathbf{p}}^2+i\delta).
\end{aligned}
\end{equation}

By power counting, second-order self-energy diagrams have one more power of gas parameters than the first-order.  In the dilute region, the mean-field energy is of the same order as LHY energy, i.e. $ \delta g \sim g'\phi' $, and the two gas parameters $\delta \phi$ and $\phi'$ are not of the same order.  Therefore to obtain the correction to the phonon spectrum, we only need to consider second-order diagrams of $\Sigma_{cc}$ due to interaction with spin excitations which is of the order $ g'n\phi' $. As shown in Fig.~\ref{fig1}(B), these diagrams consists of coupling constant $g'$ and first-order Green's function $G_d(p)$ or $\hat{G}_d(p)$.  Physically, it indicates the correction to the phonon spectrum comes from interaction between phonon and spin excitations.  All other higher-order effects are negligible.  We obtain the corrected self-energy $\Sigma_{cc}$ given by
\begin{equation}\label{2nd-selfenergy}
\begin{aligned}
\Sigma_{cc}^{11}(0)&=2n\delta g+\frac{80}{3\sqrt{\pi}}ng'\sqrt{na'^3},\\
\Sigma_{cc}^{20}(0)&=n\delta g+\frac{16}{\sqrt{\pi}}ng'\sqrt{na'^3},
\end{aligned}
\end{equation}
and $\mu=n{\delta g}+\dfrac{32}{3\sqrt{\pi}}ng'\sqrt{n a'^3}$ satisfying the gapless condition $ \mu=\Sigma_{cc}^{11}(0)-\Sigma_{cc}^{20}(0) $.  With these corrections, the phonon Green's function near the pole is approximately given by
\begin{equation}\label{G_c}
\begin{aligned}
G_c(p)&=(1-\lambda)\frac{p^0+\epsilon^0_\mathbf{p}+\Sigma_{cc}^{11}(-p)-\mu_c}{(p^0+\mathcal{E}^c_\mathbf{p}-i\delta)(p^0-\mathcal{E}^c_\mathbf{p}+i\delta)},\\
\hat{G}_c(p)&=\frac{-(1-\lambda)\Sigma_{cc}^{22}(p)}{(p^0+\mathcal{E}^c_\mathbf{p}-i\delta)(p^0-\mathcal{E}^c_\mathbf{p}+i\delta)},
\end{aligned}
\end{equation}
where $ \lambda=8\sqrt{n a'^3/\pi}\ll 1 $.
From the pole, we obtain the phonon energy
\begin{equation}\label{pole}
\mathcal{E}^c_\mathbf{p}=\sqrt{\epsilon^0_\mathbf{p}\left(\epsilon^0_\mathbf{p}+2n\delta g+32n g'\sqrt{n a'^3}/\sqrt{\pi}\right)}.
\end{equation}
The ground state energy is given by $ E_0/V=\int \mu dn $. Then using the zero-pressure condition $ P=\mu n-E_0/V=0 $, we obtain the equilibrium density for droplet $ n=25\pi\gamma^2 a'^{-3}/4096 $, or equivalently, $ \delta g=-64g'\sqrt{na'^3}/(5\sqrt{\pi}) $.
In the long-wavelength limit, the phonon energy is linearly dispersed $ \mathcal{E}^c_\mathbf{p}\approx\hbar vp$, where the phonon velocity is given by $v=\sqrt{-\delta g n/(4m)}$.  This positive phonon velocity shows that the phonon mode is stablized by the interaction between spin and phonon excitations.  The healing length of the phonon is given by $\xi_c=\hbar/(\sqrt{2}mv)=\sqrt{2\hbar^2/(-\delta gnm)}$, much larger than that of the spin excitation $\xi_d=\sqrt{\hbar^2/(2g'nm)}$ by a factor of the order of $\sqrt{\phi'}$.

In Eq. \eqref{pole}, the imaginary part of the pole is absent, indicating that Beliaev-damping rate of a phonon is greatly suppressed and much smaller than that of a spin excitation. This conclusion can be also drawn by considering the process as follows.  A phonon with energy $ \mathcal{E}^c_\mathbf{p} $ cannot be split into two excitations with at least one spin excitation due to energy and momentum conservation, i.e. $ \mathcal{E}^c_\mathbf{p}<\mathcal{E}^c_\mathbf{q}+\mathcal{E}^d_{\mathbf{p-q}} < \mathcal{E}^d_\mathbf{q}+\mathcal{E}^d_{\mathbf{p-q}} $ at finite $\mathbf{p}$.  The Beliaev damping only happens when a phonon at finite $\mathbf{p}$ decays into two phonons, with damping rate the order of $p^5 \phi'^3 $.  In comparison, the damping of spin excitations is more complicated. In the lowest order, a spin excitation with energy $ \mathcal{E}^d_\mathbf{p}\approx\hbar p\sqrt{g'n/m} $ can only decay into a spin excitation with energy $ \mathcal{E}^d_\mathbf{q}$ and a phonon with energy $\mathcal{E}^c_\mathbf{{p-q}} $.  Note that in the case with asymmetric intraspecies interactions $ a_{\alpha\alpha}\ne a_{\beta\beta} $, there is an additional on-shell process, allowing spin excitation decaying to two other spin excitations.

\begin{figure}[]
	\includegraphics[width=\columnwidth]{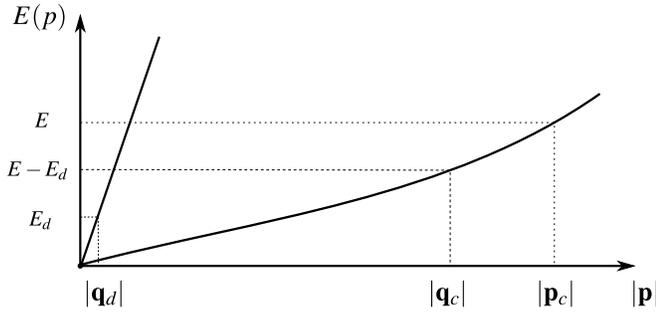}
	\caption{\label{fig2} Excitation spectrum of a quantum droplet. If a phonon with wavevector $ |\textbf{p}|_c\lesssim 1/\xi_c $ and energy $E$ decays to a spin excitation with $ |\textbf{q}_d| $ and $E_d$, and another phonon with $ |\textbf{q}_c| $ and $E-E_d$, the momentum conservation will be violated.  The phonon is undamped up to second order in $\phi'$. }
\end{figure}

\textit{Asymmetric case}--
The intraspecies interactions of the $^{39}\mathrm{K}$ quantum droplet in the experiment are asymmetric.  Our above method can be readily generalized to this case.  For simplicity we assume $\langle \hat{\alpha}_0 \rangle>0$ and $\langle \hat{\beta}_0 \rangle>0$.  We can perform the following unitary transformation to decouple the two components in the Bogoliubov Hamiltonian,
\begin{equation}\label{transformation}
\begin{pmatrix}
\hat{c}_\mathbf{p}\\
\hat{d}_\mathbf{p}\\
\end{pmatrix}
=
\begin{pmatrix}
\sqrt{n_\alpha/n} & \sqrt{n_\beta/n}\\
\sqrt{n_\beta/n} & -\sqrt{n_\alpha/n}
\end{pmatrix}
\begin{pmatrix}
\hat{\alpha}_\mathbf{p}\\
\hat{\beta}_\mathbf{p}\\
\end{pmatrix}.
\end{equation}
In the new representation  the expectation values of new annihilation operators are given by $\langle \hat{c}_0 \rangle=\sqrt{nV}$ and $\langle \hat{d}_0 \rangle=0$. As the symmetric case, the Bogoliubov quasi-particle of type-$c$ is the phonon of density fluctuation, and that of type-$d$ corresponds to spin fluctuation which changes the density ratio of $\alpha$ and $\beta$ components.  The phonon excitation energy given by the Bogoliubov theory is imaginary in the long-wavelength limit, and our approach is essentially the same as before.  We look for the second-order contribution to the self-energy $\Sigma_{cc}$ due to the interaction between phonon and spin excitations.  We obtain the same results as given in Eq.~\eqref{2nd-selfenergy}-\eqref{pole}, except that in the asymmetric case the coupling constants in these equations are given by
\begin{equation}
	\begin{aligned}
		\delta g =&-\gamma\sqrt{g_{\alpha\alpha}g_{\beta\beta}}/(1-\gamma),\\
		g'= &\sqrt{g_{\alpha\alpha}g_{\beta\beta}}/(1-\gamma).
	\end{aligned}
\end{equation}

In the long wavelength limit, the phonon excitation energy is given by $ \mathcal{E}^c_\mathbf{p}\approx\hbar vp$, where the phonon velocity is
\begin{equation}\label{velocity}
 v=\dfrac{5\pi\hbar\gamma^{3/2}(1-\gamma)}{64m\sqrt{a_{\alpha\alpha}a_{\beta\beta}}}\approx\dfrac{5\pi\hbar\gamma^{3/2}}{64m\sqrt{a_{\alpha\alpha}a_{\beta\beta}}}.
\end{equation}
Details of this derivation are given in the supplemental material~\cite{supplementary}.

The ground-state energy of the quantum droplet can be also computed in this Green's function approach and the leading correction to the mean-field energy is consistent with LYH energy in Ref \cite{petrovQuantumMechanicalStabilization2015}.   From this ground-state energy and zero-pressure condition, we obtain the positive compressibility given by $ {\kappa}=-4/(\delta g n^2) $.
The sound velocity obtained from thermodynamic relation $ v=1/\sqrt{mn\kappa} $ agrees with the phonon behavior in Eq.\eqref{pole}, which indicates that phonons of the quantum droplet is in the superfluid hydrodynamic region.

\textit{Discussion}--
The phonon excitation is associated with the propagation of density sound wave. For the $ {}^{39}\mathrm{K} $ droplet in the experiment \cite{cabreraQuantumLiquidDroplets2018}, the sound velocity according to Eq.\eqref{velocity} is about $ 7.8\times10^{-4}\text{m/s} $. It is possible in experiments to measure sound velocity in droplet. For example, one can excite density perturbations and observe the propagation of sound waves as the single component system \cite{andrewsPropagationSoundBoseEinstein1997}. Another method is stirring a droplet with fixed velocity \cite{ramanEvidenceCriticalVelocity1999}. According to Landau's criterion, the condensate is dissipationless when scan velocity is below the critical velocity, i.e. the sound velocity. Alternatively, a Bragg spectroscopy can be used to determine the excitation spectrum \cite{steinhauerExcitationSpectrumBoseEinstein2002}.

Measuring the sound velocity requires a droplet as large  as possible. On the length scale $ \xi $, the interaction energy becomes comparable with the kinetic energy \cite{pethickBoseEinsteinCondensation2008}.  The healing length obtained from Eq.\eqref{pole} is given by
$\xi_c=64\sqrt{a_{\alpha\alpha}a_{\beta\beta}}\gamma^{-3/2}/(5\sqrt{2})
$.  A homogeneous droplet ball with diameter $ \xi_c $ contains about $ N_\xi=\dfrac{32}{15\pi}\gamma^{-5/2} $ atoms. For $ {}^{39}\mathrm{K} $ droplet mixture \cite{cabreraQuantumLiquidDroplets2018}, the number $ N_\xi $ is about 4400. When the atom number $ N\gg N_\xi $, the excitations of droplet are collective, which helps to observe sound velocity. In comparison, a droplet with atom number $ N\lesssim N_\xi $ may facilitate global excitations under a local disturbance \cite{cheineyBrightSolitonQuantum2018}.

Unlike a single component BEC with attractive interaction \cite{pethickBoseEinsteinCondensation2008}, a mixture BEC with negative mean-field energy can be long lived even at large atom number. This indicates the unstable-miscible boundary is modified due to interaction between two quasi-particle branches. The actual boundary is closer to the zero-pressure condition $ \gamma=\dfrac{64}{5\sqrt{\pi}}\sqrt{na'^{3/2}} $, rather than $ \sqrt{a_{\alpha\alpha}a_{\beta\beta}}+a_{\alpha\beta}=0 $.  Thus our results are readily applicable to trapped BEC mixture with small positive mean-field energy and fixed density ratio $ n_{1}/n_{2}\approx\sqrt{a_{22}/a_{11}} $ in the miscible region.

\textit{Conclusion}--
In conclusion, we go beyond the Bogoliubov theory to study excitations in a binary quantum droplet.  Different from the predictions of the Bogoliubov theory, the phonon excitation is found to be stable with a positive sound velocity, which can be readily tested in experiments.  The Beliaev damping of phonon is greatly suppressed in a quantum droplet.  Our results indicates that the phonon of the quantum droplet is in superfluid hydrodynamic region.

We would like to thank Z.-Q. Yu and B. Liu for helpful discussions. This work is supported by the National Basic Research Program of China under Grant No. 2016YFA0301501.

\textit{Note added.}-- Recently, we became aware of two related theoretical papers \cite{huConsistentTheorySelfbound2020} and \cite{otaLeeHuangYangDescriptionSelfbound2020} that give other perspectives of the phonon stability.


%

\end{document}